# Nucleation Pathways in Barium Silicate Glasses

Matthew E. McKenzie[1*], Binghui Deng[1], D. C. Van Hoesen[2], Xinsheng Xia[3], David E. Baker[1], Aram Rezikyan[1], Randall E. Youngman[1], and K. F. Kelton[2,3]

[1]Science and Technology Division, Corning Research and Development Corporation, Corning, New York 14831, USA

[2]Department of Physics, Washington University, St. Louis, MO 63130, USA

[3]Institute of Materials Science and Engineering, Washington University, St. Louis, MO 63130, USA

*Correspondence to mckenzieme@corning.com

## Abstract

Nucleation is generally viewed as a structural fluctuation that passes a critical size to eventually become a stable emerging new phase. However, this concept leaves out many details, such as changes in cluster composition and competing pathways to the new phase. In this work, both experimental and computer modeling studies are used to understand the cluster composition and pathways. Monte Carlo and molecular dynamics approaches are used to analyze the thermodynamic and kinetic contributions to the nucleation landscape in barium silicate glasses. Experimental techniques examine the resulting polycrystals that form. Both the modeling and experimental data indicate that a silica rich core plays a dominant role in the nucleation process.

## Introduction

Nucleation and crystallization in glasses is a rich subject area covering tunable materials for high technology applications to fundamental scientific inquires [1] [2] [3] [4] [5] [6]. The final nucleated glass, or glass-ceramic, may have improved material properties, such as toughness [6], over the parent glass. By understanding the nucleation event, it may be possible to design an improved glass-ceramic or recognize how the devitrification process occurs, which is crucial for the glass making industry. The Classical Nucleation Theory (CNT) [1] [7] [8] [9] is most commonly used to describe the nucleation and growth of the emerging new phases. Qualitatively, CNT provides a good framework to understand the process. However, problems arise when this method is used to obtain a quantitative description (thermodynamic and kinetic) of the nucleation phenomenon. In this work, we focus on homogeneous nucleation rates that occur in the glass far below the melting temperature ($\Delta T = T_m - T$, where $T_m$ is the melting temperature). At these temperatures, the critical nuclei are only 1 to 2 nanometers in size, making it difficult to experimentally probe their size evolution.

According to CNT, the thermodynamic properties of the evolving new crystalline phase are considered size independent and equivalent to the properties of the macroscopic phase, with a sharp interface separating the nucleating cluster from the parent phase (i.e., the capillarity approximation). For instance, in CNT the interfacial free energy between a crystal cluster and the glass will be the same for very small clusters as it is for macroscopic ones. Assuming this, the work of cluster formation, $W_n$, in CNT is a function of cluster size ($n$) having a radius $r$ and a constant interfacial free energy term $\sigma$, as shown in Eq. (1).

$$W_n = 4\pi r^2 \sigma - \frac{4}{3}\pi r^3 \Delta g \tag{1}$$

The driving free energy for nucleation and growth can be computed using the difference of the Gibbs free energies of the bulk crystal and the amorphous phase, $\Delta g$. The critical cluster size, $n^*$, is the size below which the clusters are, on average, dissolving and above which they are, on average, growing.

The assumption of a sharp interface between the cluster and the parent phase is a reasonable one for the case of vapor-liquid nucleation that CNT was originally developed to address. However, in dense phase systems where the parent phase is in intimate contact with the developing new phase, the interface cannot be as sharp. This was originally proposed by Turnbull to explain differences between his experimental results and predictions of the CNT for nucleation in liquid mercury [10]. To better describe the emerging cluster and interface region, two theories have been developed: the Diffuse Interface Theory (DIT) and Density Functional Theories (DFT). DIT does not assume the capillarity approximation, but instead describes the work in terms of an assumed profile of the Gibbs free energy as a function of distance from the cluster center [11] [12] [13]. The DFT theories [14] [15] [16] are based on an order parameter description of the phase transition. Computer modeling of the nucleation process has generated great insight. Within supercooled water, immobile regions form volumes where nucleation can occur [17]. Like the pathways of the emergent phase, one can ask how does the interface develop?

A long-standing problem of using CNT for glass-ceramic systems is the work of critical cluster formation, $W_n$, as a function of temperature. Based on experimental studies at low temperatures (lower than the maximum nucleation temperature, $T_{max}$), $W_n$ begins to increase with decreasing temperature, rather than continuing to decrease as expected [18] [19] [20] [21] [22]. This discrepancy has had at least five different possible explanations. These include: (1) a change in the sign of the temperature dependence for the interfacial free energy with a temperature dependence [10], (2) a change in the driving free energy with temperature [21], (3) a cross-over in the kinetics of long-range diffusion and interfacial attachment [23], (4) a decrease in the thermodynamic driving free energy due to elastic stress [21], and (5) a variation of the cooperatively rearranging regions of the critical nuclei that may account for kinetics and longer annealing times [24]. However, these explanations alone have not been able to clarify the description of nucleation at such deep undercoolings (50 degrees below $T_{max}$). A common thread in these explanations is the thermodynamic landscape as a function of temperature, which is explored in this work.

By illuminating the thermodynamic nucleation contributions as a function of chemistry and temperature, one may be able to detect differences in the nucleating pathways for the sub-critical nuclei. Considering a pathway change, the growing cluster composition is important as it may elucidate other routes (i.e., micro-phase separation leading to nucleation). Over recent years, a "two-step" nucleation mechanism has been gaining attention, where a disordered droplet will form first, followed by crystallization [4] [25] [26] [27] [25].  Probing these intermediate clusters is important in identifying these paths. How distinct are these two steps of nucleation? Can they occur concurrently?  Most nucleation theories assume that the clusters grow by *monomer* attachment and detachment. This is a straightforward counting method for the nucleation of droplets of water in vapor, where the monomer is well defined. But, how well does this scheme work in glass-crystal nucleation?  Does one count neutrally charged oxide units (every $i$-$SiO_2$ or $i$-BaO); does one track individual elements; or would one count by unit-cells (every $i$-{5BaO 8SiO_2} or $i$-{1BaO 2SiO_2} units)? The barium silicate system is known to contain multiple crystalline polymorphs within the same crystal [28], so how would monomer tracking look in this case? How would the evolution of the nuclei modify the interfacial region and resulting instantaneous interfacial free energy? To best explore this nucleation cluster free energy landscape, we choose to track the cluster evolution by every oxide unit. Between CNT and DIT, experiment and models, we will probe the nucleation pathways. Reality is more complicated and multiple concurrent intermediate phases will be shown to play a role when the final state is reached [4] [25].

*Classical Nucleation Theory background*

Gibbs [29] [30] studied phase transitions that occur under near equilibrium conditions, where the probability of fluctuations that lead to the new stable phase is small. This corresponds to a large nucleation barrier.  He found that the barrier decreases when the system is quenched more deeply into the metastable region. Later, Volmer and Weber [31] constructed CNT based on Gibb's reversible work for a critical cluster and a new kinetic model that they introduced.  In this formulation, the work of formation, $W(n)$, of a cluster of the new phase containing $n$ particles, is used to describe an equilibrium cluster size distribution for clusters smaller than the critical size.  The critical cluster size, $n^*$, or denoted by the cluster radius, $r^*$, corresponds to the maximum work of formation $W^*$ (the asterisk in the superscript denote a critical size value).  Assuming a steady-state cluster distribution, Becker and Döring [32] showed that the steady-state nucleation rate, *I*, is proportional to the Boltzmann weighted probability of the critical fluctuation and a kinetic prefactor, *A\**,

$$I = A^* \exp\left(-\frac{W^*}{k_B T}\right), \qquad (2)$$

where *T* is temperature and $k_B$ is the Boltzmann constant. Turnbull and Fisher extended this nucleation formulation to include crystallization from a liquid by relating the prefactor to kinetic processes in the liquid [33]. For deep undercooling, the influence of the kinetic prefactor is difficult to assess. The key kinetic term in the prefactor is the diffusion rate, which for crystallization from liquids and glasses is typically determined from the viscosity, assuming the Stokes-Einstein relation. However, the viscosity is difficult to measure and model at low temperature. Also, it is believed that there is a breakdown of the Stokes-Einstein relation at these temperatures [34] [35].

*Diffuse Interface Theory background*

Based on experimental nucleation studies, Turnbull argued that instead of having a sharp interface between the parent and new phase, as assumed by Gibbs, the interface was actually somewhat diffuse. Gránásy [11] [12] and Spaepen [13] independently proposed a phenomenological model to account for the changing thermodynamic quantities through the diffuse interface. Here, the work of cluster formation, *W*, is a function of the Gibbs free energy, *g*,

$$W = \int_0^\infty 4\pi r^2 g(r)\, dr \qquad (3)$$

The Gibbs free energy at radius *r* is *g(r)* = Δ*h* - *T*Δ*s*, where Δ*h* and Δ*s* are the enthalpy and the entropy difference between the glass and crystal respectively. Within the DIT, there is an offset in the thermodynamic quantities, Δ*h* and Δ*s,* as a function of distance from the crystal cluster center, *r*. While the precise profile for *g(r)* through the interface is unknown, by approximating Δ*h(r)* and Δ*s(r)* as a series of step functions Eq. 3 can be directly solved, assuming that the Gibbs free energy between the bulk glass and the center of the crystal is known (i.e. from the Turnbull approximation or from specific heat data of the glass and crystal).

## Results

This section will discuss (1) the results of the GCMC model for barium silicate nucleation, (2) a comparison to the DIT and ordering between the solid and liquid phases, (3) low temperature concerns, and (4) the experimental results. Different polymorphs of the barium silicate glass are considered: $1BaO \cdot 2SiO_2$, $3BaO \cdot 5SiO_2$, $5BaO \cdot 8SiO_2$, and $4BaO \cdot 6SiO_2$. For convenience, these different crystals will be referenced by their barium silicate stoichiometry. For example, $5BaO \cdot 8SiO_2$ will be labeled 5-8.

### Barium silicate GCMC study

Nucleation modeling is sensitive to the force field used. It was found that the melting temperature for the 1-2 was 60 K higher than the experimental value (1693 K). This is reflected in the nucleation temperatures in the model, where it is predicted that the maximum nucleation temperature, $T_{max}$, occurs at 1090 K, higher than the experimental value (985 K [22]). Following CNT it was found that the maximum work per crystal polymorph occurred between 2 to 3 unit cells (4 to 6 unit cells for the smallest 1-2 system). The maximum work of formation (i.e. the work at the critical size) for the 1-2 and 5-8 glasses in shown in Fig. 1. This is in good agreement with previous experimental results [22], where the maximum work at 1090K for the 1-2 and 5-8 glasses are 32 $k_BT$ (vs. the experimental value of 31 $k_BT$) and 24 $k_BT$ (28 $k_BT$ experiment) respectively. Interestingly, from Fig. 1, there is a continued decrease in the work of formation. This lower temperature had to be sampled 30% longer than the other temperatures due to slow convergence. This was caused by a slowly evolving cluster structure with very cohesive elements (i.e., the force field is too attractive).

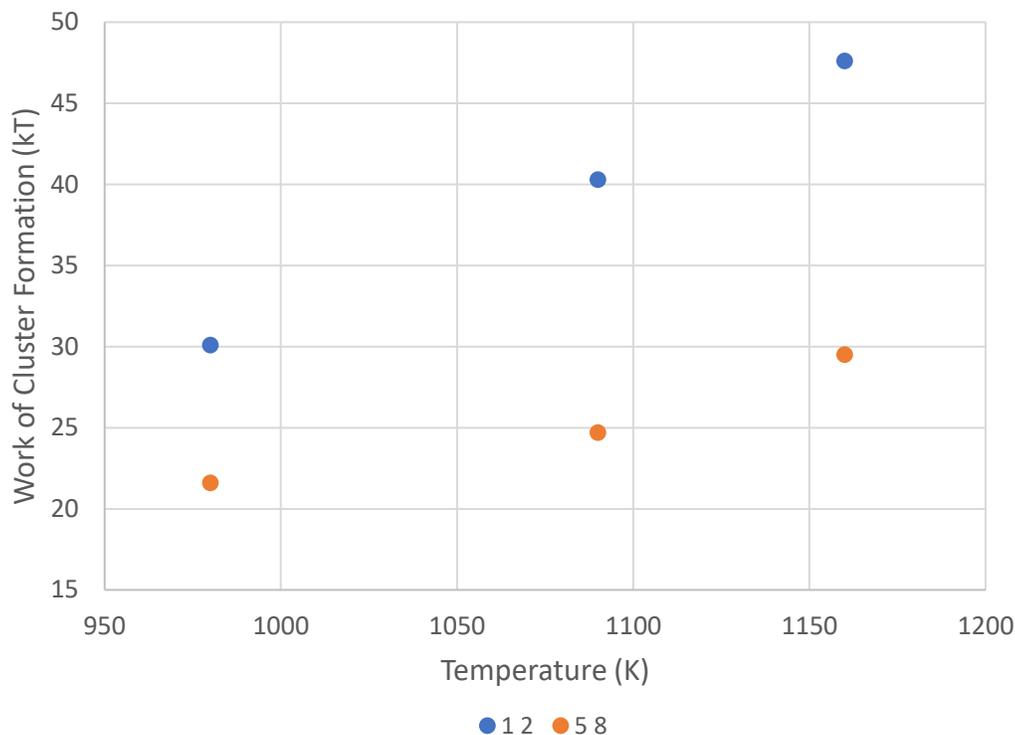

Figure 1.  The work of cluster formation in barium silicate as function of temperature for the 1-2 and 5-8 glasses.

While the overall melt stoichiometry may be that of a barium silicate crystal, it is heterogeneous and there is a possibility that a region may contain a higher than average concentration of barium (or silica).  As shown in Fig. 2, this will guide where and how the nucleation occurs. Note that $W^*$ in Fig. 1 is the work of the cluster (Eqs. 7-9) at the first maximum in the Nucleation Free Energy (NFE) contour map shown in Fig. 2. The NFE map is created by thermodynamic integration of the cluster's work.  This Fig. 2 only has the thermodynamic contributions as there are no continuum diffusion kinetics in the GCMC model. According to CNT, one must follow the diagonal line representing the polymorph stoichiometry. For all polymorphs and at all three temperatures, there is a protected valley labelled as **A**. The first unit cell of each of these polymorphs occur in this region (except, it is 4 unit cells of the 1-2 system). These pre-critical clusters then must climb the NFE barrier region (marked with **B** in Fig. 2) to reach more stable cluster configurations.

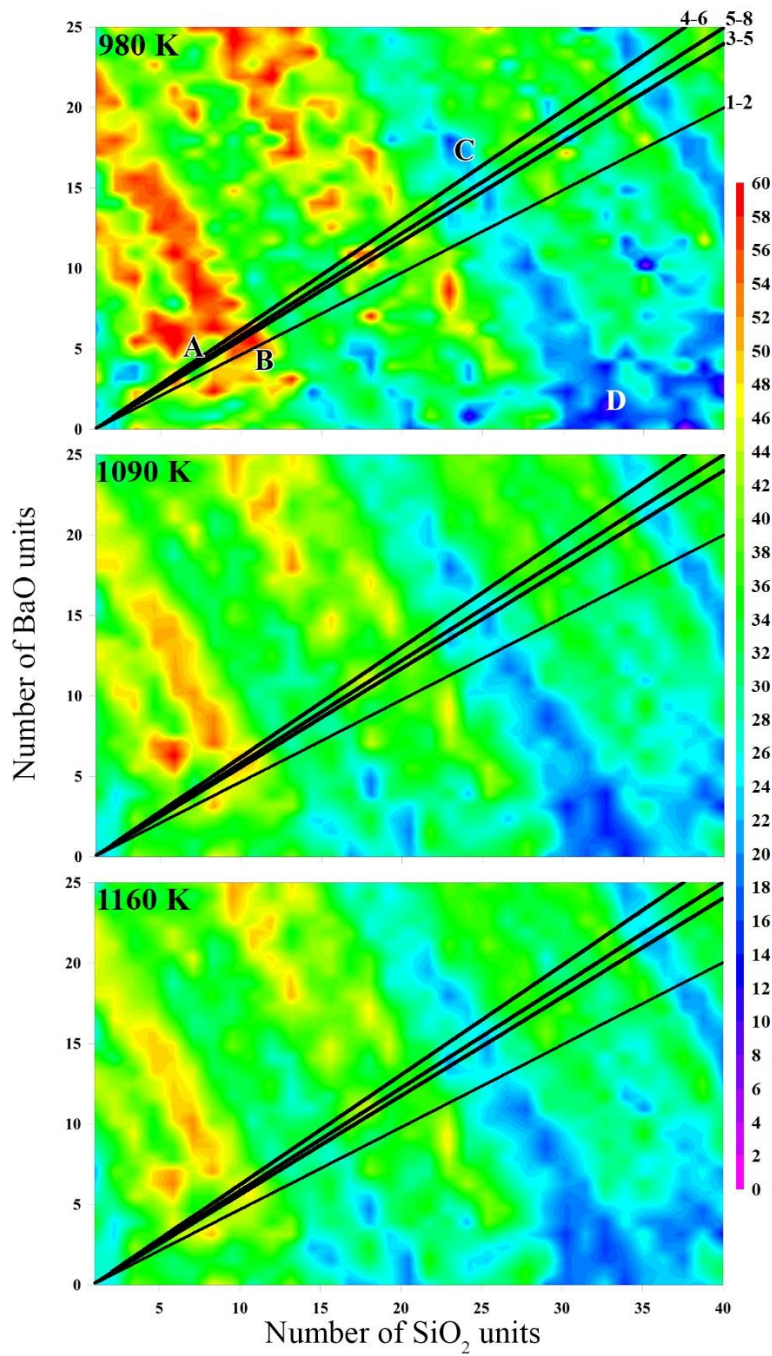

Figure 2. NFE contour maps of the sampled barium silicate clusters. The number of BaO and SiO$_2$ units are on the *y* and *x* axis, respectively. The temperature is in the upper left of each contour plot. The NFE contour coloring scheme is located on the right, in units of k$_B$T. Different barium silicate

polymorphs correspond to the different diagonal lines in each of these three maps and their designation is labelled in the upper left side of the 980K map. The labels are explained in the text and appear in the 980K plot.

Comparing the 980, 1090, and 1160 K contour plots in Fig. 2, the NFE landscape becomes smoother with increasing temperature, and the size of the higher energy regions (red contours) decreases. A decreasing NFE barrier at higher temperatures will cause the nucleation rate to increase. However, as the CNT polymorph stoichiometry lines are traversed across Fig. 2, different paths may be taken to reach the next cluster size (e.g., first add $SiO_2$ then BaO or *vice versa*). No matter which route is taken to go beyond this large NFE barrier in the 980K case, it is necessary to either travel through it or dissolve BaO units to form a more silica rich cluster. With increasing temperature, more routes become accessible since the NFE contours decrease.

Beyond this barrier at **B**, the clusters will reach another NFE minimum at **C**. This blue region extends diagonally towards the silica axis, marked by **D**. The system should follow the lowest free energy path. With such a low NFE region at **D**, a glass heterogeneity of this composition, which is quite distinct from that of the bulk, may in fact serve as the first step in the nucleation process. Given a specific bulk barium silicate stoichiometry, is it possible to have regions that are enriched in silica and deficient in barium? Glass melting research has been working on reducing compositional inhomogeneities for a long time and this remains a concern for glass manufacturing [36] [37] [38] [39] [40] [41] [42]. However, it is possible to find such regions within the glass structure. Depending on the raw material, how the glass is melted (re-melted), and the annealing cycles used, all may affect the compositional inhomogeneity. These factors would also affect how the cooperatively rearranging regions in the glass melt evolve [43] [44]. Analyzing compositional variation in glass melts, however, is outside the scope of this work, though is a critical aspect of our ongoing experimental studies of nucleation in the barium silicate system.

The ordered structure should have comparable amounts of baria (BaO) and silica ($SiO_2$) if it grows according to CNT throughout the cluster volume. Here, we analyze the last 1,000 Monte Carlo configurations by tracking the amount of silica to the total sum of silica and baria found in the inner 50% of the cluster volume. The standard deviation of this analysis is less than the resulting contour intervals. For clusters smaller than 10 silica and 10 baria the inner volume is very small and numerical artifacts are seen. Surprisingly, for all three temperatures shown, the inner volume is silica rich (Fig. 3). Using the diagonal line as the 5-8 unit cell line (the other polymorphs would fall nearby on either side), it appears nucleation seems to be mainly driven by a silica rich environment. This compares favorably with the study by Avramov et. al. [45] on topological constraint theory [46] [47] and nucleation where they found that the network needs to be slightly rigid to enable crystallization. The silica structures would be more rigid than the barium oxide network modifier structures, simply due to a higher degree of network connectivity provided by fewer non-bridging oxygens and more fully polymerized silicate tetrahedra. Additionally, the formation of silica (e.g. cristobalite) would be the most favored structure on this NFE landscape (Fig.2 labelled **D**). Finding cristobalite or quartz during the glass making process is common and undesired (a.k.a. devitrification).

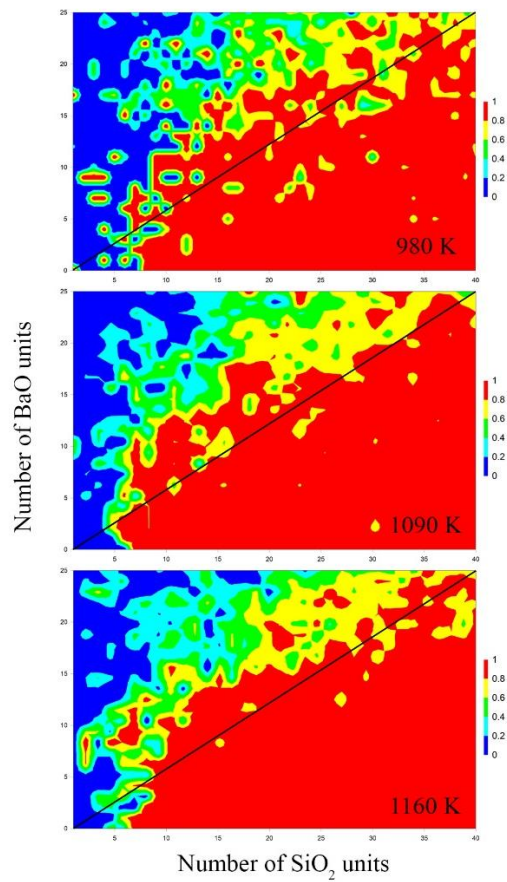

Figure 3. Spatial composition of the silica to total oxide ratio for all clusters studied. Temperatures are shown on the lower right side of the contours.

## Comparison to DIT and ordering between phases

Clearly, CNT has its deficiencies, especially in describing the atomic environment of the pre-critical clusters. DIT brings in a more detailed thermodynamic description of the order parameter as a function of distance through the cluster and into the melt. Here, we used a molecular dynamics simulation where we explicitly model both the crystal cluster and the melt (Fig. 4). In DIT the critical cluster size is determined in the usual way, by setting the derivative of the work of cluster formation at a given temperature equal to zero and solving for the radius (or number of particles/monomers). Here, DIT predicts that the critical cluster size has a radius ≈12 Å at 980K, which is slightly larger than the biggest clusters modeled with the GCMC approach (radius ~10 Å). The critical size is smaller from the GCMC approach compared to DIT. It is reasonable considering the NFE landscape has hills and valleys (fluctuations) that will lead to a critical nucleus. The same fluctuations were seen in a previous lithium silicate study [27] as the cluster grows. Additionally, the NFE landscape in Fig. 2 seems to become smoother at larger sizes. This qualitatively compares well to the MD potential energy results in Fig. 5 (discussed below). To further illuminate these fluctuations and the developing crystal-glass interface, we studied a critical size nuclei (r = 12 Å) and a post-critical one (r = 41 Å).

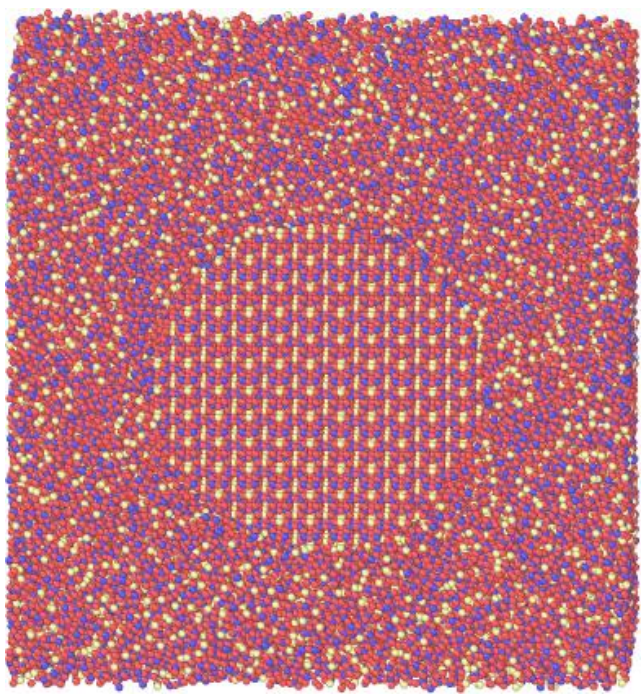

Figure 4. Starting snapshot of the combined crystal and melt of the 1-2 system at T=1040K. The radius of this cluster is 41 Å. The silicon, oxygen, and barium atoms are colored red, yellow, and blue, respectively.

Using the MD glass-ceramic model, we radially averaged the potential energy and bond-orientational order parameter Q4 (Fig. 5). Since this is an average over a spherical radius, depending on the bin it may cross over different crystal planes – the Q4 parameter is especially sensitive to this. The Potential Energy of a crystal is equivalent to the enthalpy, assuming that no pressure-volume work is being performed. For DIT's critical cluster of 12 Å, there are highs and lows of energy corresponding with crystal lattice positions. The magnitude and pattern clarity continue into the melt (right side of dotted line in Fig. 5(a)).  There is a decrease in the Q4 structure when passing through this interface (Fig. 5(b)). The nearest melt layer shows signs of the Q4 and Potential Energy patterning of the crystal. This shows that within the simulation, the cluster is stable and the MD model agrees with DIT's prediction. Interestingly, there is little difference between the crystal and the melt in Fig. 5 (a) & (b) (left vs. right sides), meaning that the cluster is not too dissimilar to the glass.  This indicates that the cluster is more like the interface than a crystalline cluster with a sharp interface.  It is worth mentioning these values extend into the melt, inferring a larger interface, or fluctuation.

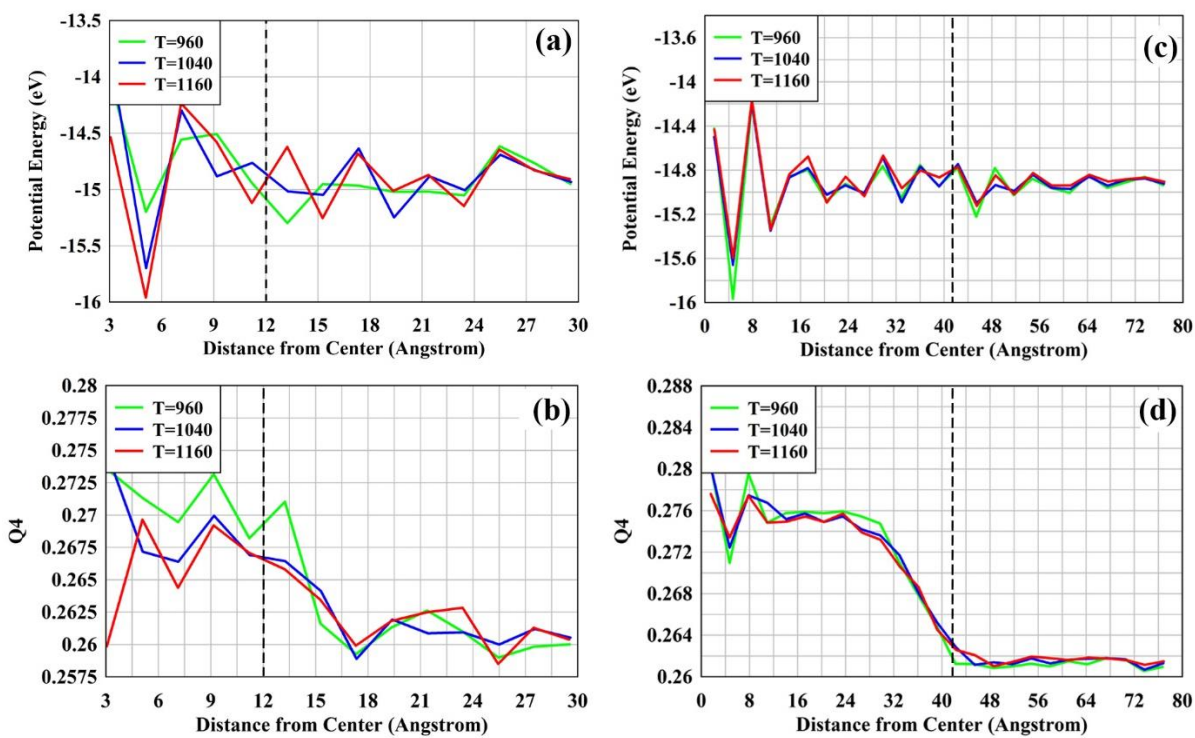

Figure 5. Potential Energy and bond-orientational order parameter Q4 as a function of distance from the center for a cluster with radius 12 (panels a & b) and 41 Å (panels c & d). The vertical dotted line represents the cluster radius (the right side of the dotted line is the melt). The bin size for the 12 and 41 Å clusters are 2 and 3 Å respectively. The first couple of points in each of these plots have numerical artifacts from the binning process.

The post critical cluster (radius =41 Å, Fig.5 panels c & d) shows oscillations of the Potential Energy that follow the crystalline ordering; the magnitude is lower in the melt (being randomly distributed). Additionally, the larger crystal cluster shows a clear loss of order. This ordering begins to fall off around 30 Å. In other words, the cluster has an interfacial width of approximately 10 Å! As the cluster grows, the interface becomes sharper and less significant relative to the cluster radius, being more consistent with the assumptions of CNT. This finding is in line with previous studies [48] [49]. Interestingly, the crystal's oscillating Potential Energy extends into the melt (~10Å, Fig. 5 (c)) further than the bond order does. This may indicate that the correct layering occurs first, followed by an increasing density (lowering of entropy). Then, the crystalline ordering occurs to reduce the total volume. This is opposite of Tanaka's [25] work, where the crystallization and growth occur with a density fluctuation followed by the crystalline bond ordering. However, this work agrees with a metallic liquid nucleation study [50] where an icosahedral ordering occurs first, followed by crystallization. Clearly, there are nucleation nuances that remain open.

**Lower temperature concerns**

From Fig. 5, there is no temperature effect on the width of interfacial region in the atomistic simulation. One might expect the interfacial width to widen as temperature increases. Additionally, the lowest temperature in the GCMC study had slow convergence since it samples a flatter configuration energy space. To better address this concern, a simple and straightforward investigation of the barium silicate diffusion coefficient allows one to assess the force field at lower temperatures and gives insight into the nucleation kinetics.

The Ba diffusion is shown in Fig. 6; Ba diffuses faster than O and Si. To obtain the diffusion coefficient a slope of unity must be reached on a plot of Mean Squared Displacement (MSD) vs time. If the slope is less than unity the atoms are trapped in a cage and the diffusion coefficient cannot be measured until they escape. At only the two highest temperatures has the slope been attained for a proper calculation of the diffusion coefficient. The lines in Fig. 6 are composed of three restarted simulations (of 200ns each). If one approximates a linear relationship as the slope changes from each additional run, for the T=1190K case it would require an additional 4,600 ns trajectory to reach a slope of unity. That is assuming a linear relationship, which is unlikely because these cooler temperatures become nearer to the glass transition temperature, it is more of a measure of the bare minimum time needed. It was found that the diffusion coefficient for Ba was 0.01378 nm$^2$/ns and 0.14563 nm$^2$/ns at temperature of 1500 and 1773 K, respectively. This value is two to four orders of magnitude *faster* than an estimation made using the crystal growth velocity [51] , where it was extrapolated to be 7.94×10$^{-4}$ nm$^2$/ns at T=1693K. Clearly, the modeled diffusion coefficient is faster than the experiment suggests. Predicting lower temperature diffusion would suffer from similar inaccuracies and differences in the glass transition temperature. The difficulties in the GCMC thermodynamic model can be effectively mitigated with some additional sampling, however the dynamic contributions would be orders of magnitude more time consuming.

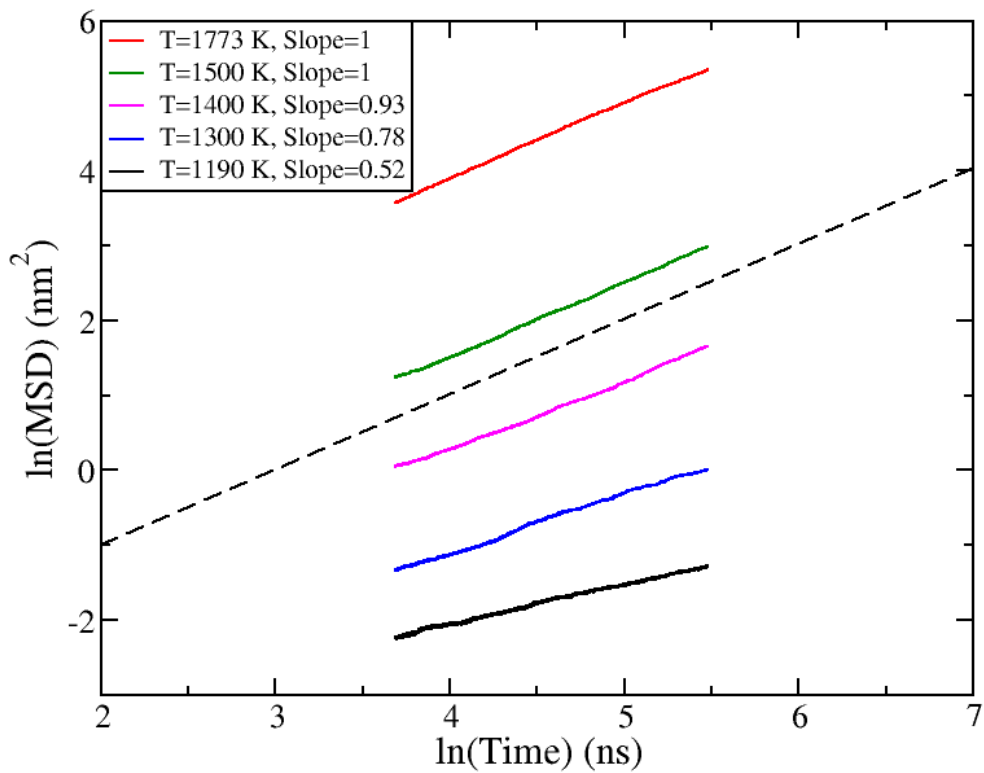

Figure 6. Barium natural log of the Mean Squared Displacement (MSD) vs. natural log of time plot for temperatures ranging from 1190 to 1773 K. The colored lines are composed of three sequential simulation restarts.

**Experimental Study: 5-8 system**

Shown in Fig. 7 are SEM images of the 5-8 composition after a nucleation treatment for 2 hours at 780°C. These images compare the morphology with and without acid etching, showing large changes when briefly exposed to the 0.1% Hydrofluoric acid (HF) solution (etchant). There appears to be a center region in the fractured and not etched sample (Fig. 7). The acid etching generates voids and sharpens the resolution of the nuclei center region versus the outer region of the spheroidal regions. Silica is especially susceptible to HF acid, suggesting that an explanation for these differences is that the cores of these nuclei are Si-rich and thus more readily removed during etching. To look for such compositional differences, additional SEM and EDS analyses were conducted on the nucleated 5-8 glass, the results of which are given in Fig. 8. Indeed, the EDS analysis of the center region versus the outer parts of the spheroids confirms a significant compositional difference, with the core region being enriched in Si and depleted in Ba. This is certainly consistent with the observation of different etching behavior across the nucleus. Further, the results of the GCMC study also found that all clusters at any temperature contain a silica rich core (Fig. 3) and that a silica rich region might serve as a better nucleation site (Fig. 2, label **D**). There is a size discrepancy between the model and the SEM image, where the modeled cluster radii towards the upper left corner in Fig. 2 are ~12-20Å, while the radii of the clusters shown in Fig. 7 and 8 are 600nm. In spite of these size differences, good agreement between the predictions of the model and the experimental data is observed for compositional fluctuations within the 5-8 nuclei.

Fractured and etched

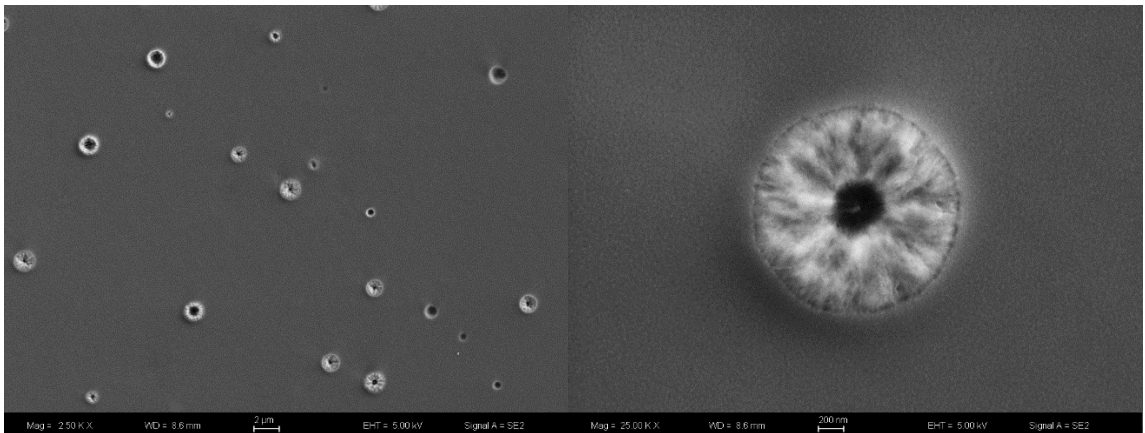

Fractured and not etched

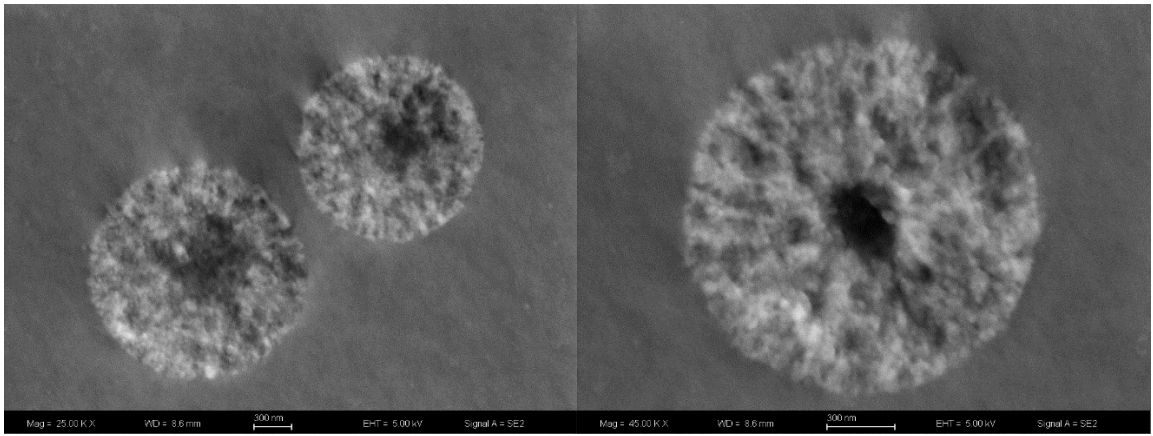

Figure 7. SEM images of the 5-8 glass sample nucleated at 780°C for two hours. The upper two side by side images are for the fractured and etched surface. The lower two images are the fractured and not etched surface.

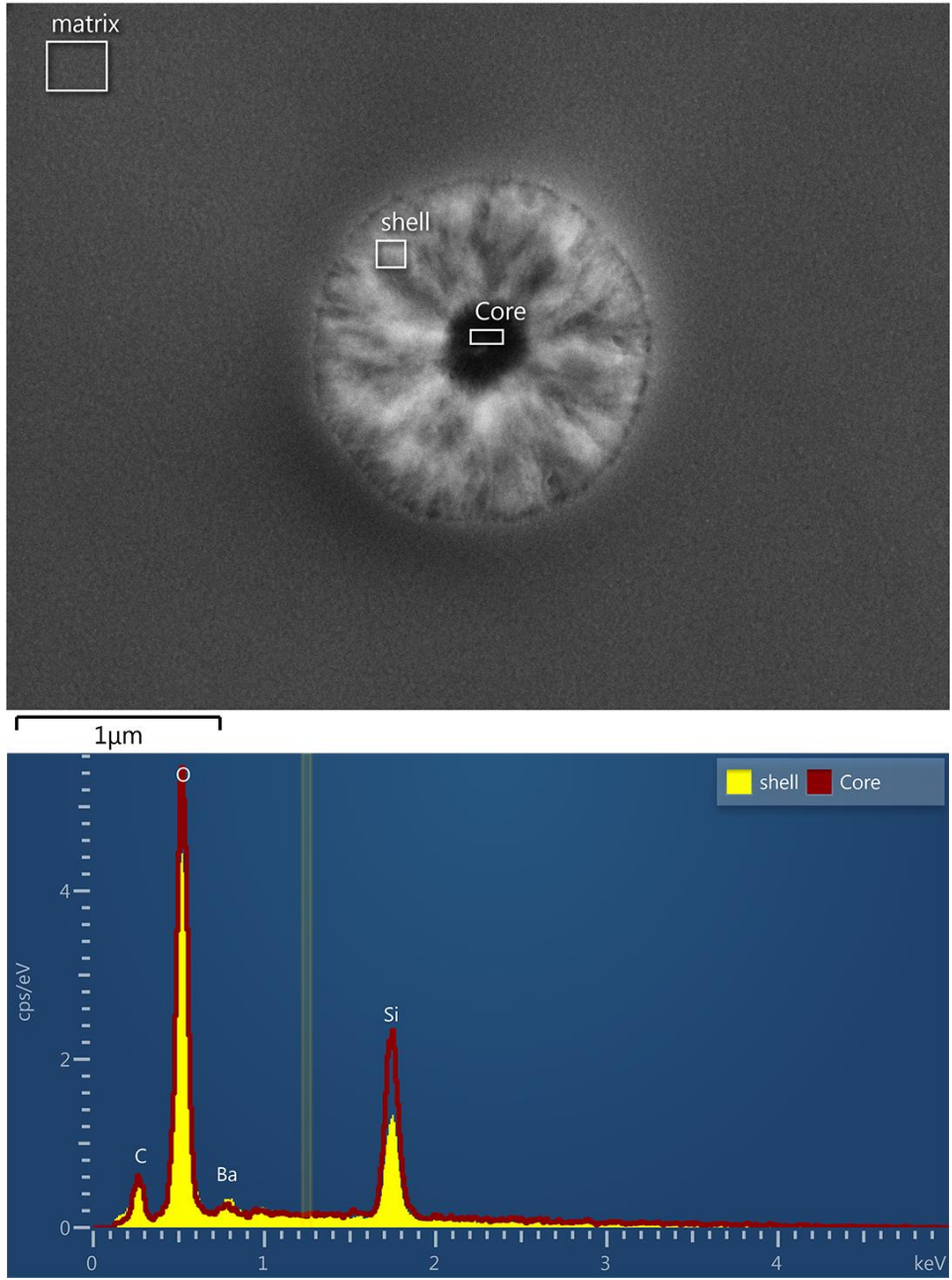

Figure 8. EDS of the core and shell of the fractured and 0.1% HF etched sample.

**Experimental Study: 1-2 system**

The 1-2 system was nucleated at 850 °C for 30 minutes. Other samples were also nucleated at lower temperatures and longer times and showed similar crystalline features. The morphology of the 1-2 crystal is needle-like and distinct from that of the 5-8's spheroid shape. These needle-like crystals also exhibit the same core and shell organization as the 5-8 (Fig. 9 (a)). A closer inspection using Bright Field STEM shows a variety of crystal types and twinning planes. The 1-2 unit cell line in Fig. 2 is further away from the other barium silicate crystal lines in the same figure. Yet, a similar polycrystalline sample emerges (Fig. 9) in that we find needle-like and spherulite crystals having 1-2, 5-8, and 2-3 compositions. This occurs due to evolving local composition (and chemical potential) and the ability to transverse to nearby CNT crystal polymorph pathways (i.e., there are no large barriers that are parallel to the different barium silicate crystal line in Fig. 3).

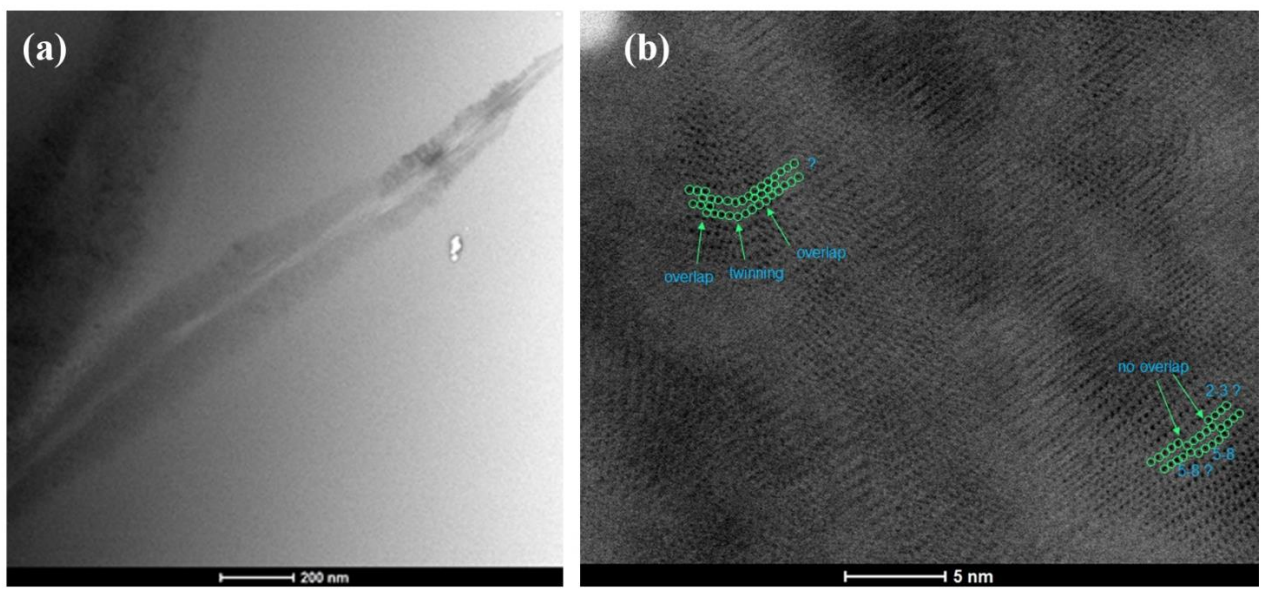

Figure 9. Bright-field STEM images of the 1-2 sample at different magnifications (see scale bar lower middle of each panel). Suggested crystal stoichiometries are highlighted in (b). Question marks indicate uncertainty about the stoichiometry because the it is not clear if rows of atoms at the nanocrystal edge terminate fully capturing the whole row for a given stoichiometry.

**Discussion & Future Work**

In this paper, we demonstrate the richness of nucleation by: (1) showing how a nuclei develops through CNT, DIT, different modeling efforts, and experiment; (2) identifying the feasibility of an alternative silica rich pathway; (3) identifying different nearby CNT pathways leading to co-crystallization of other Ba-Si forms, and (4) demonstrating that the thermodynamic contributions to nucleation also decrease and cannot describe the apparent $T_{max}$ anomaly. The Classical Nucleation Theory discretizes the macroscopic world into a microscopic realm, but, these results show that nucleation is a complex process having a greater depth than CNT ever envisioned. The discretization is, in fact, a spectrum of possible microstates all interacting with one another, as shown in the GCMC model. The Diffuse Interface Theory begins to fill in the spectrum by adding an interfacial region and connecting the emerging cluster to the melt; as the cluster grows the interfacial width decreases. The modeling shows that for sub-critical and critical cluster sizes, the interface is a significant portion of the cluster. The modeled work of cluster formation as a function of cluster composition can lead to a variety of nucleation pathways, where each cluster has a different thermodynamic driving force and interfacial free energy, all dependent on the local composition. This is shown from the experimental investigation, where independently from overall melt stoichiometry, other barium silicate crystals are identified.

An interesting behavior, common between the different pathways, is the silica rich core. Experimental measurements have shown that the inner core of the 5-8 cluster is silica rich, which qualitatively agrees with predictions from the GCMC model. If the CNT discretized lines are not followed, but the whole barium silicate landscape is followed, the low NFE region on the silica rich axis, having low barium content (Fig. 2, label D), may serve as the first step in the nucleation process. Qualitatively, this agrees well with studies of nucleation in supercooled water, where immobile regions allow water molecules to nucleate. In this case, the silica rich regions of a melt are more thermodynamically stable and less mobile (being a network former), allowing barium silicate to nucleate from them. A scientific question then arises, being a silica rich heterogeneity, would this be a result of a poorly mixed glass occurring via heterogeneous nucleation instead? This may be the guiding principle of cluster growth and devitrification.

The number of different pathways as a function of temperature and composition tends to decrease as a function of temperature, shown by the more pitted and hillier NFE landscape in Fig. 2 as temperature decreases. However, the work of formation for the critical cluster does tend to decrease with decreasing temperature, unlike what was found in previous studies [21] [24] [10] [23]. An attempt at the kinetic contribution (diffusion coefficient) failed due to force field limitations. Therefore, this study cannot explain the low temperature anomaly in the work of formation, except that the thermodynamic part has been confirmed. This work does illustrate the possible competing pathways that are available during a nucleation event. The work of formation is dependent on the nucleating pathways which makes Eq. 1 inherently more complex. As a consequence, Volmer and Weber's kinetic model [31] would have different activation energies for adjoining clusters based on their thermodynamic reactant state shown here.

Even for a simple binary glass, nucleation is very complex and additional studies are needed. Once scientific investigation answers a question, it unlocks many more questions. For instance, does silica clustering drive glass-ceramic nucleation? Do all polymorphs of barium silicate exist in other compositions? What does the core tell us about nucleation? How does the silica rich core develop as a function of time and temperature? Can the core's creation be perturbed by adding a fast diffusing alkali metal to the system? Can a phase sensitive glass force field be generated? Further work into advancing Diffuse Interface Theory and Density Functional Theory methods, such as competing pathways, regions of slow mobility, and interface development with advancements in experimental methodologies, will increase our understanding of nucleation.

Acknowledgements

© 2020 Corning Incorporated. All Rights Reserved.

We thank Drs. Bruce G. Aitken, David R. Heine, Ling Cai, Bryan R. Wheaton, and Troy D. Loeffler for their thought-provoking discussions.

## Computational Methods

The results presented here are based on studies from a Grand Canonical Monte Carlo (GCMC) model, molecular dynamics (MD) models, and experimental observations. The GCMC model is used to explore the nucleation free energy (NFE) landscape of cluster sizes and compositions, a technique that has shown to give good agreement with experimental results [27]. MD is used to probe the crystal-melt interface, which is an atomistic analogue to DIT. One item the GCMC model does not account for is diffusion, so the MD model is used to investigate this contribution. Experimentally, the nucleated barium silicate glasses were analyzed using Scanning Electron Microscopy (SEM), Transmission Electron Microscopy (TEM), and Energy Dispersive Spectroscopy (EDS). This section outlines each of these models, analysis techniques, and the experimental methods used.

## Grand Canonical Monte Carlo (GCMC) Model

This model [27] [52] was developed to study nucleation in a lithium disilicate glass. The work of cluster formation is not described by the usual surface tension and thermodynamic driving terms but instead tracks (1) the amount of work that is required to form a cluster, (2) the energy required to crystallize that cluster and (3) the associated solvation energy. The model only simulates the cluster; it does not simulate the atoms in the glass. This allows the computation to focus solely on the cluster development instead of using computational resources to move atoms in the glass region that do not directly partake in the nucleation event. Each of the clusters experiences the continuum field of the glass in the form of an implicit solvation model. For completeness some details of the simulation technique are discussed below. Full details of the computer code and methods can be found in Refs. [53] [54] [55] [56] [57].

<u>Cluster Formation</u>

This model uses the Grand Canonical ($\mu VT$) ensemble, where the clusters can have a fluctuating number of atoms. The Grand Canonical ensemble has the cluster physically separated from (but thermodynamically coupled to) the liquid phase. This is similar to the Gibbs approach except it does not explicitly model the liquid phase. We only track the development and growth of one cluster in the Grand Canonical ensemble. It is difficult to directly obtain the chemical potential, $\mu$, for these glass-ceramic systems, whether one uses a model approach (e.g., Widom particle insertion technique [58]) or experiment. Instead, from statistical mechanics, one can derive the chemical potential to be equivalent to the product of the Boltzmann constant, absolute temperature, and the natural logarithm of the number density [59]. This constant volume term in the ensemble is enforced by an energy-based Stillinger-type cluster criterion. The cluster is defined as a group of atoms in which every atom has at least one neighbor with interaction energy equal to or greater than 50% of the Si-O bond energy (226 kJ/mol). The atomistic models use the force field developed by Pedone *et al*. [60]. All clusters from 1-40 $SiO_2$ units and 0-25 BaO units have been sampled with at least 800 million Monte Carlo cycles for temperatures of 980K, 1090K, and 1160K (representing below, at, and above $T_{max}$).

We utilize the Aggregation-Volume-Bias Monte Carlo (AVBMC) move type [53] to bypass the low acceptance rate of a traditional Metropolis Monte Carlo move [35] in these simulations. An AVBMC move has two options: (1) cluster addition (out →in), or (2) cluster subtraction (in→out). The target particle *j* is then randomly selected in the cluster. If this is a cluster addition move, the code will randomly select a particle from the ideal glass phase and place it inside the $V_{in}$ volume near particle *j*. In the case of a subtraction move, a random particle within the $V_{in}$ region of particle *j* is chosen as a candidate for deletion. The energy change between the current configuration *A* and the trial configuration *B* (based on addition or subtraction) is then calculated, i.e., $\Delta E = E_B - E_A$. Within the $V_{in}$ region of the cluster $N_{in}$ is the total number of particles (sums to *N*-1, where *N* is the total particles of the cluster, as *j* is excluded). The trial move is accepted with the following probabilities:

1. Cluster addition

$$acc(A \rightarrow B) = \min\left[1, \frac{V_{in} \times N \times \exp\left(\frac{\mu}{k_B T}\right) \times \exp\left(-\frac{\Delta E}{k_B T}\right)}{(N_{in}+1) \times (N+1)}\right] \quad (4)$$

2. Cluster destruction

$$acc(A \rightarrow B) = \min\left[1, \frac{N_{in} \times N \times \exp\left(-\frac{\mu}{k_B T}\right) \times \exp\left(-\frac{\Delta E}{k_B T}\right)}{V_{in} \times (N-1)}\right] \quad (5)$$

Incorporating the energetic and entropic factors experienced in the associating and non-associating volumes will lead to a higher probability for accepting a trial move, hence increasing the efficiency of the Monte Carlo moves and leading to an effective evolution of the cluster distributions from pre- to post-critical sizes. The Monte Carlo move types are chosen at a frequency of 30% AVBMC and 70% translational movements. The resulting cluster structure and calculated energies are invariant on the Monte Carlo move frequencies, however the total simulation length does depend on the move ratio. It was found that an increased amount of translational movement tends to help the cluster re-organize whenever a particle is added/removed.

Crystal Transition

Steinhardt order parameters [9] [61] were used to measure the amount of energy required to move the atoms in the amorphous structure into the crystalline atomic positions. This technique has been used for liquid to solid and solid to solid phase transitions [62] [63] [64]. At the heart of this computational technique, the bond vectors and angles are calculated and compared to reference crystal values. The differences in the order parameter values can be sampled to calculate the amount of energy required to form a crystal from the initial amorphous positions [27].

Solvation energy

The solvation energy is found using the Implicit Glass Model (IGM) [65]. IGM is a member of the Generalized Born implicit solvation model [66] [67] family that are normally used for proteins. This is a per atom solvation term that is incorporated into the cluster's total energy. The parameters used in this work came from ref. [65]. While this term typically only adds ~1 k$_B$T unit to the work of cluster formation, it is very important since the sampling of the cluster formation is conducted only on one cluster in vacuum (and no residual glass).

Nucleation Free Energy (NFE) Analysis

AVBMC's purpose is to increase the frequency of cluster growth and destruction event. This is still insufficient when attempting to sample clusters near the critical size. Due to the low occurrence probability of clusters at the critical size, a typical barrier is of the order 30 k$_B$T; this means that one must sample $e^{30}$ times to observe the event! This is solved by using an umbrella sampling scheme [68] in which a bias potential is developed during the simulation to enhance the sampling frequencies. The bias potential is removed in the nucleation free energy analysis.

In the Grand Canonical ensemble, the probability, $P$, of observing a particular state described by $n$ particles and their configuration $r^n$ in a biased distribution is

$$P(n, r^n) = \exp(\mu n \beta) \exp(-\beta\, E(N, r^n) B(n)), \qquad (6)$$

where $\beta$ is 1/k$_B$$T$ (where k$_B$ is the Boltzmann constant and $T$ is the temperature in absolute units and $B(n)$ is the biasing potential as a function of cluster size $n$. The bias potential is multiplied by the AVBMC acceptance criteria (Eq. 4 and 5) as $W(n+1)/W(n)$ (for addition) or $W(n-1)/W(n)$ (for cluster destruction). The probability of having clusters of size $n$ in the bias distribution, $P(n)_{biased}$ is computed; the probability of the unbiased distribution is then given by

$$P(n)_{unbiased} = \frac{P(n)_{biased}}{W(n)} \qquad (7)$$

The objective is to choose a proper weighting function, to have a statistically precise estimate of $P(n)_{biased}$ with the least amount of computational time. A reasonable criterion is to have an equal probability of observing each cluster size $n$. This can also provide the maximum possible sampling of each size for a given simulation length. The choice of this weighing function is





$$W(n) = \frac{1}{P(n)_{unbiased}} \quad (8)$$

This gives an equal probability to sample all cluster sizes of *n* in the biased distribution. The free energy of cluster formation with this biasing potential becomes

$$\Delta G_i = -k_B T \ln(W(n)) - k_B T \ln\left(\frac{P(n)}{P(1)}\right) \quad (9)$$

Unfortunately, this choice generates a new problem, as the unbiased probabilities are not known (and their determination is the objective of the simulation). There are different ways to obtain the appropriate biasing potential. From knowledge of this algorithm, there are two reasonable starting choices: (1) conduct a few million samplings of unbiased small cluster sizes to initiate the bias, or (2) if another bias potential is known (either generated at another temperature or extrapolated from the previous point) those values can be used. Since this bias approach is an iterative and adaptive process, many millions of Monte Carlo cycles are needed to ensure convergence. Convergence is reached when the bias potential does not change by more than 1% of its value. After every 10,000 Monte Carlo cycles the bias potential is re-calculated. In total, each cluster size requires only one CPU core for approximately 2-3 weeks of clock time. It should be mentioned that for the lowest temperature studied (980 K), this convergence required 30% more Monte Carlo cycles than for the other temperatures.

**Molecular Dynamics**

The Monte Carlo model allows one to probe the complete nucleation pathway. A molecular dynamics model is used to study the explicit interface between the crystal and the glass, the evolving melt structure near the interface, and the bulk diffusion of the barium silicate glass. A stoichiometric barium disilicate glass was first generated by randomly inserting 250,000 atoms into a cubic simulation box and equilibrating them at 3000 K for 2 ns in the Canonical (NVT) ensemble. This liquid was then continuously cooled down to 300 K over a period of 6 ns in the Isothermal-isobaric (NPT) ensemble under atmosphere pressure. The final sample dimension was 23.2 nm × 46.3 nm × 2.89 nm. Afterwards, a spherical void with a radius of 4.1 nm was created in the center of the glass by deleting atoms inside the region and refilling it with a piece of pre-made spherical BaO 2SiO$_2$ crystal of the same radius. The premade crystal was cut out of a large piece of crystal generated using the crystal structure information reported in ref [69] [70]. Because of the atom deletion and insertion operations, the system usually lost charge balance. To remedy this, additional atoms in the glass were randomly chosen and deleted to maintain charge neutrality for the final sample. The bulk diffusion study used the same sized simulation box and heat treatment, but without the inserted crystal.

To monitor the evolution of the glass-crystal interface, the sample was gradually heated up from 300 K to the target temperature (960 K, 1040 K, 1190 K) in the period of 1 ns using the NPT ensemble, and further held at the target temperature for 1 ns. The bond-orientational order parameter Q4, introduced by Steinhardt et al. [61], was calculated to characterize the local orientational order in the atomic structure. This parameter was expected to capture the structural differences between the glass, the interface, and the crystal. To obtain a better statistical Q4 value, the atoms were spatially binned into spherical shells from the center of the simulation box to the edge, using a bin size of 0.3 nm.

These simulations were conducted using LAMMPS [71] with a timestep of 2 fs. Temperature and pressure were well controlled via the Nose-Hoover [72] [73] thermostat and barostat, respectively. Periodic boundary conditions (PBC) were applied in all directions over the course of these operations. The force field created by Pedone et. al. [60] was used to capture the short-range interactions for both phases. For the long-range columbic interactions, the damped shifted force (dsf) method, with a cutoff of 8 Å and damping parameter of 0.25 Å$^{-1}$, was used to speed up calculations.

**Experimental Methods**

TEM images were acquired using a bright-field detector in an aberration corrected FEI Titan high resolution transmission electron microscope (HRTEM) operated at 200 keV electron beam energy in the scanning mode (STEM). The samples were prepared by the conventional TEM preparation method (mechanical polishing and dimpling followed by Ar-ion polishing).

Samples were prepared for SEM analyses by fracturing and acid etching in a 0.1% HF solution for 10 seconds to help reveal surface topography. An amorphous conductive carbon coating was evaporated onto the samples to reduce charging. Secondary electron images were acquired in a Zeiss 1550 field emission scanning electron microscope at an accelerating potential of 5 keV and 200 picoamps beam current. Energy-dispersive X-ray spectroscopy analyses were acquired using an Oxford Instruments 80 mm$^2$ silicon drift detector and AZtec x-ray microanalysis software.

## References

[1] K. Kelton and A. Greer, Nucleation in Condensed Matter, vol. 15, Pergamon Oxford, 2010.

[2] E. D. Zanotto, "A bright future for glass-ceramics," *Am. Ceram. Soc. Bulletin,* vol. 89, no. 8, pp. 19-27, 2010.

[3] Q. Fu, G. Beall and C. Smith, "Nature-inspired design of strong, tough glass-ceramics," *MRS Bulletin,* vol. 42, no. 3, pp. 220-225, 2017.

[4] G. C. Sosso, J. Chen, S. J. Cox., M. Fitzner, P. Pedevilla, A. Zen and A. Michaelides, "Crystal Nucleation in Liquids: Open Questions and Future Challenges in Molecular Dynamics Simulations," *Chem. Rev.,* vol. 116, pp. 7078-7116, 2016.

[5] J. Deubener and W. Holand, "Editorial: nucleation and crystallization of glasses and glass-ceramics," *Front. Mater.,* vol. 4, p. 14, 2017.

[6] W. Holand and G. H. Beall, "Applications of Glass-Ceramics," in *Glass-Ceramic Technology*, Hoboken, NJ, John Wiley & Sons, Inc., 2012.

[7] P. F. James, "Kinetics of crystal nucleation in silicate glasses," *J. Non-Cryst. Solids,* vol. 73, pp. 517-540, 1985.

[8] I. Gutzow, D. Kashchiev and I. Avramov, *J. Non-Cryst. Solids,* vol. 73, pp. 477-499, 1985.

[9] S. Auer and D. Frenkel, "Numerical Simulation of Crystal Nucleation in Colloids," in *Advanced Computer Simulation. Advances in Polymer Science*, vol. 173, Springer, Berlin, Heidelberg , 2005, pp. 149-208.

[10] D. Turnbull, "Kinetics of Solidification of Supercooled Liquid Mercury Droplets," *J. Chem. Phys.,* vol. 20, p. 411, 1952.

[11] L. Gránásy, "Diffuse Interface Approach to Vapor Condensation," *Europhys. Lett.,* vol. 24, no. 2, pp. 121-126, 1993.

[12] L. Gránásy, "Diffuse interface theory of nucleation," *J. Non-Cryst. Solids,* vol. 162, no. 3, pp. 301-303, 1993.

[13] F. Spaepen, "Homogeneous nucleation and the temperature dependence of the crystal melt interfacial tension," *Solid State Physics,* vol. 47, pp. 1-32, 1994.

[14] D. W. Oxtoby and R. Evans, "Nonclassical nucleation theory for the gas–liquid transition," *J. Chem. Phys.,* vol. 89, pp. 7521-7530, 1988.

[15] J. F. Lutsko, "Recent developments in classical density functional theory," *Adv. Chem. Phys.,* vol. 144, p. 1, 2010.

[16] R. Evans, "The nature of the liquid-vapour interface and other topics in the statistical mechanics of non-uniform, classical fluids," *Adv. Phys.,* vol. 28, pp. 143-200, 1979.

[17] M. Fitzner, G. C. Sosso, S. J. Cox and A. Michaelides, "Ice is born in low-mobility regions of supercooled liquid water," *PNAS,* vol. 116, no. 6, pp. 2009-2014, 2019.

[18] J. Schmelzer, O. Potapov, V. Fokin, R. Muller and S. Reinsch, *J. Non-Cryst. Solids,* vol. 333, pp. 150-160, 2004.

[19] V. M. Fokin, E. D. Zanotto, J. Schmelzer and O. Potapov, *J. Non-Cryst. Solids,* vol. 351, pp. 1491-1499, 2005.

[20] V. M. Fokin and E. D. Zanotto, "Continuous compositional changes of crystal and liquid during crystallization of a sodium calcium silicate glass," *J. Non-Cry. Solids ,* vol. 353, no. 24-25, pp. 2459-2468, 2007.

[21] A. S. Abyzov, V. M. Fokin, A. M. Rodrigues, E. D. Zanotto and J. W. P. Schmelzer, "The effect of elastic stresses on the thermodynamic barrier for crystal nucleation," *J. Non-Cryst. Solids,* vol. 432, pp. 325-333, 2016.

[22] X. Xia, D. C. Van Hoesen, M. E. McKenzie, R. E. Youngman, O. Gulbiten and K. F. Kelton, "Time-dependent nucleation rate measurements in BaO·2SiO2 and 5BaO·8SiO2 glasses," *J. Non-crystal. Solids,* vol. 525, p. 119575, 2019.

[23] K. F. Kelton, "Time-dependent nucleation in partitioning transformations," *Acta Materialia,* vol. 48, no. 8, pp. 1967-1980, 2000.

[24] V. M. Fokin, A. S. Abyzov, E. D. Zanotto, D. R. Cassar, A. M. Rodrigues and J. W. P. Schmelzer, "Crystal nucleation in glass-forming liquids: Variation of the size of the "structural units" with temperature," *J. Non-Cryst. Solids,* vol. 447, pp. 35-44, 2016.

[25] H. Tanaka, "Bond orientational ordering in a metastable supercooled liquid: a shadow of crystallization and liquid–liquid transition," *J. Stat. Mech.,* p. P12001, 2010.

[26] J. F. Lutsko, "How crystals form: A theory of nucleation pathways," *Sci. Adv.,* vol. 5, no. 4, 2019.

[27] M. E. McKenzie and J. C. Mauro, "Hybrid Monte Carlo technique for modeling of crystal nucleation and application to lithium disilicate glass-ceramics," *Comput. Mater. Sci.,* vol. 149, pp. 202-207, 2018.

[28] M. Czank and P. R. Buseck, "Crystal chemistry of silica-rich barium silicates," *Zeitschrift fur Kristallographie ,* vol. 153, pp. 19-32, 1980.

[29] J. W. Gibbs, "On the equilibirum of heterogeneous substances," *Trans. Connect. Acad.,* vol. 3, pp. 108-248, 1878.

[30] J. W. Gibbs, "On the equilibirum of heterogeneous substances," *Trans. Connect. Acad.,* vol. 3, pp. 343-524, 1878.

[31] M. Volmer and A. Weber, "Nuclei formation in supersaturated states (transl.)," *Z. Phys. Chem.,* vol. 119, pp. 227-301, 1926.

[32] V. R. Becker and W. Doring, "Kinetic treatment of grain-formation in super-saturated vapours," *Ann. Physik,* vol. 24, pp. 719-752, 1935.

[33] D. Turnbull and J. C. Fisher, "Rate of nucleation in condensed phase systems," *J. Phys. Chem.,* vol. 17, pp. 71-73, 1949.

[34] S. Pan, Z. W. Wu, W. H. Wang, M. Z. Li and L. Xu, "Structural origin of fractional Stokes-Einstein relation in glass-forming liquids," *Scientific Reports,* vol. 7, p. 39938, 2017.

[35] S. Wei, Z. Evenson, M. Stolpe, P. Lucas and C. A. Angell, "Breakdown of the Stokes-Einstein relation above the melting temperature in a liquid phase-change material," *Sci. Adv.,* vol. 4, no. 11, 2018.

[36] N. L. Bowen, "THE IDENTIFICATION OF "STONES" IN GLASS.," vol. 1, no. 9, pp. 594-605, 1918.

[37] H. Le Chatelier, *Bull. Soc. Min. Fr.,* vol. 150, no. 39, 1916.

[38] J. W. Creig, "Immiscibility in silicate melts; Part II," *Am. J. Sci.,* vol. 13, no. 5, pp. 133-154, 1927.

[39] P. F. James, "Liquid-phase separation in glass-forming systems," *J. Mater. Sci.,* vol. 10, no. 10, pp. 1802-1825, 1975.

[40] C. Cammarota, A. Cavagna, I. Giardina, G. Gradenigo, T. S. Grigera, G. Parisi and P. Verrocchio, "Phase-Separation Perspective on Dynamic Heterogeneities in Glass-Forming Liquids," *Phys. Rev. Lett.,* vol. 105, p. 055703, 2010.

[41] G. W. Morey, The properties of glass, 2nd edn., New York: Reinhold Publishing, 1954.

[42] J. E. Shelby, Introduction to Glass Science and Technology, Cambridge, UK: The Royal Society of Chemistry, 1997.

[43] G. Adam and J. H. Gibbs, "On the Temperature Dependence of Cooperative Relaxation Properties in Glass-Forming Liquids," *J. Chem. Phys.,* vol. 43, no. 139, 1965.

[44] K. L. Ngai and R. W. Rendell, "Couplings between the cooperatively rearranging regions of the Adam–Gibbs theory of relaxations in glass-forming liquids," *J. Chem. Phys.,* vol. 94, no. 3018, 1991.

[45] I. Avramov, R. Keding, C. Russel and R. Kranold, "Precipitate particle size distribution in rigid and floppy networks," *J. Non-Crystal. Solids,* vol. 278, no. 1-3, pp. 13-18, 2000.

[46] J. F. Phillips, *J. Non-Crysta. Solids,* vol. 34, p. 153, 1979.

[47] M. F. Thorpe, *J. Non-Crystal. Solids,* vol. 57, p. 355, 1983.

[48] D. W. Oxtoby and P. R. Harrowell, "The effect of density change on crystal growth rates from the melt," *J. Chem. Phys.,* vol. 96, pp. 3834-3843, 1992.

[49] C. K. Bagdassarian and D. W. Oxtoby, "Crystal nucleation and growth from the undercooled liquid: A nonclassical piecewise parabolic free-energy model," *J. Chem. Phys.,* vol. 100, pp. 2139-2148, 1994.

[50] K. F. Kelton, G. W. Lee, A. K. Gangopadhyay, R. W. Hyers, T. J. Rathz, J. R. Rogers, M. B. Robinson and D. S. Robinson, "First X-Ray Scattering Studies on Electrostatically Levitated Metallic Liquids: Demonstrated Influence of Local Icosahedral Order on the Nucleation Barrier," *Phys. Rev. Lett.,* vol. 90, p. 195504, 2003.

[51] D. C. Van Hoesen, X. Xia, M. E. McKenzie and K. F. Kelton, "Modeling nonisothermal crystallization in a BaO 2SiO2 glass," *J. Am. Ceram. Soc.,* vol. 103, pp. 2471 - 2482, 2019.

[52] T. Loeffler, "github.com," 2019. [Online]. Available: https://github.com/mrnucleation/ClassyMC. [Accessed 1 Jan Jan].

[53] B. Chen, J. I. Siepmann and M. Klein, "Simulating Vapor-Liquid Nucleation of Water: A Combined Histogram-Reweighting and Aggregation-Volume-Bias Monte Carlo Investigation for Fixed-Charge and Polarizable Models," *J. Phys. Chem. A,* vol. 109, pp. 1137-1145, 2005.

[54] B. Chen, J. I. Siepmann, K. J. Oh and M. L. Klein, "Aggregation-volume-bias Monte Carlo simulations of vapor-liquid nucleation barriers for Lennard-Jonesium," *J. Chem. Phys.,* vol. 115, no. 23, pp. 10903-10913, 2001.

[55] R. B. Nellas, M. E. McKenzie and B. Chen, "Probing the Nucleation Mechanism for the Binary n-Nonane/1-Alcohol Series with Atomistic Simulations," *J. Phys. Chem. B,* vol. 110, pp. 18619-18628, 2006.

[56] M. E. McKenzie and B. Chen, "Unravelling the Peculiar Nucleation Mechanisms for Non-Ideal Binary Mixtures with Atomistic Simulations," *J. Phys. Chem. B,* vol. 110, pp. 3511-3516, 2006.

[57] T. D. Loeffler, A. Sepehri and B. Chen, "Improved monte carlo scheme for efficient particle transfer in heterogeneous systems in the grand canonical ensemble: application to vapor-liquid nucleation," *J. Chem. Theory Comput.,* vol. 11, pp. 4023-4032, 2015.

[58] B. Widom, "Some topics in the Theory of Fluids," *J. Chem. Phys.,* vol. 39, no. 11, pp. 2808-2812, 1963.

[59] R. Baierlein, "The elusive chemical potential," *Am. J. Phys.,* vol. 69, no. 4, pp. 423 - 434, 2001.

[60] A. Pedone, G. Malavasi, M. C. Menziana, A. N. Cormack and U. Segre, "A New Self-Consistent Empirical Interatomic Potential Model for Oxides, Silicates, and Silica-Based Glasses," *J. Phys. Chem. B.,* vol. 110, no. 24, pp. 11780-11795, 2006.

[61] P. Steinhardt, D. R. Nelson and M. Ronchetti, *Phys. Rev. B,* vol. 28, p. 784, 1983.

[62] W. Lechner and C. Dellago, "Accurate determination of crystal structures based on averaged local bond order parameters," *J. Chem. Phys.,* vol. 129, p. 114707, 2008.

[63] D. Quigley and P. M. Rodger, "A metadynamics-based approach to sampling crystallisation events," *Mol. Sim.,* vol. 35, pp. 613-623, 2009.

[64] E. E. Santiso and B. L. Trout, "A general set of order parameters for molecular crystals," *J. Chem. Phys.,* vol. 134, p. 064109, 2011.

[65] M. E. McKenzie, S. Goyal, T. Loeffler, L. Cai, I. Dutta, D. E. Baker and J. C. Mauro, "Implicit glass model for simulation of crystal nucleation for glass-ceramics," *npj Comput. Mater.,* vol. 4, no. 59, 2018.

[66] B. N. Dominy and C. L. Brooks, " Development of a Generalized Born Model Parametrization for Proteins and Nucleic Acids.," *J. Phys. Chem. B,* vol. 103, pp. 3765-3773, 1999.

[67] C. J. Cramer, Essentials of Computational Chemistry: Theories and Models, Wiley, 2004.

[68] G. M. Torrie and J. P. Valleau, *Chem. Phys. Lett.,* vol. 28, p. 578, 1974.

[69] B. Deng, J. Luo, J. T. Harris, C. S. Smith and M. E. McKenzie, "Toughening of Li2O-2SiO2 glass-ceramics induced by intriguing deformation behavior of lithium disilicate nanocrystal," *J. Amer. Ceram. Soc.,* 2019.

[70] B. Deng, J. Luo, C. M. Smith, J. T. Harris and M. E. McKenzie, "Molecular dynamics simulations on fracture toughness of Al2O3-SiO2 glass-ceramics," *Scripta Materialia,* vol. 162, pp. 277-280, 2019.

[71] S. Plimpton, "Fast Parallel Algorithms for Short-Range Molecular Dynamics," *J. Comp. Phys.,* vol. 117, pp. 1-19, 1995.

[72] S. Nosé, "A unified formulation of the constant temperature molecular dynamics methods," *J. Chem. Phys.,* vol. 81, no. 1, pp. 511-519, 1984.

[73] W. G. Hoover, "Canonical dynamics: Equilibrium phase-space distributions," *Phys. Rev. A,* vol. 31, no. 3, pp. 1695-1697, 1985.